\documentclass[conference]{IEEEtran}
\IEEEoverridecommandlockouts
\usepackage[utf8]{inputenc}
\usepackage[T1]{fontenc}
\usepackage{cite}
\usepackage{amsmath,amssymb,amsfonts}
\usepackage{algorithmic}
\usepackage{graphicx}
\usepackage{textcomp}
\usepackage[dvipsnames]{xcolor}
\usepackage{bm}
\usepackage{upgreek}
\usepackage{acro}
\usepackage{siunitx}
\usepackage{cleveref}

\usepackage{tikz}

\usepackage{pgf}
\usepackage{pgfplotstable}
\usepackage{pgfplots}
\pgfplotsset{compat=newest}

\newcommand{\veg}[1]{\bm{#1}}     
\newcommand{\mat}[1]{\bm{\mathit{#1}}} 
\renewcommand{\vec}[1]{\bm{#1}} 
\newcommand{\op}[1]{\mathcal{#1}} 
\newcommand{\vecop}[1]{\mathcal{#1}} 

\newcommand{\uv}[1]{\hat{\veg{#1}}} 
\newcommand{\matel}[1]{\begin{bmatrix} #1 \end{bmatrix}}

\newcommand{\matO}{\mathbf{0}}

\newcommand{\n}{\hat{\veg{n}}}

\newcommand{\dd}{\mathrm{d}}  

\newcommand{\jm}{\mathrm{j}}  

\newcommand{\e}{\mathrm{e}}

\newcommand{\T}{\mathrm{T}}

\DeclareMathAlphabet{\mathbbmsl}{U}{bbm}{m}{sl}

\newcommand{\matL}{\mat{\Lambda}}

\newcommand{\matS}{\mat{\Sigma}}

\newcommand{\Pl}{\mat P_{\matL}}
\newcommand{\Ps}{\mat P_{\matS}}

\newcommand{\Ev}{\veg E}

\newcommand{\Jv}{\veg j}

\newcommand{\Ts}{\vecop{T}_\text{s}}
\newcommand{\Th}{\vecop{T}_\text{h}}
\newcommand{\Top}{\vecop{T}}

\newcommand{\Iop}{\vecop{I}}

\newcommand{\vr}{\veg r}
\newcommand{\vrp}{\veg{r}^\prime}

\newcommand{\calderon}{Calder\'on}

\newcommand{\matTs}{\mat{T}_\text{s}}
\newcommand{\matTh}{\mat{T}_\text{h}}

\newcommand{\nxRWG}{\uv n \times \veg{f}}
\newcommand{\RWG}{\veg{f}}
\newcommand{\BC}{\widetilde{\veg{f}}}
\newcommand{\Patch}{p}
\newcommand{\Pyramid}{\lambda}
\newcommand{\DualPyramid}{\widetilde{\lambda}}
\newcommand{\DualPatch}{\widetilde{p}}

\newcommand{\gramPyPy}{\mat G_{\Pyramid\Pyramid}}
\newcommand{\gramPaPa}{\mat{G}_{\Patch\Patch}}
\newcommand{\gramDPyDPy}{\mat G_{\DualPyramid\DualPyramid}}

\newcommand{\LS}{\mathrm{LS}}
\newcommand{\km}{k_\mathrm{m}}

\newcommand{\pinv}{+}
\newcommand{\laplacian}{\Delta}
\newcommand{\veclaplacian}{\veg \Delta}
\newcommand{\riccatibessel}[1]{J_{#1}}
\newcommand{\riccatihankeltwo}[1]{H_{#1}^{(2)}}

\def\BibTeX{{\rm B\kern-.05em{\sc i\kern-.025em b}\kern-.08em
    T\kern-.1667em\lower.7ex\hbox{E}\kern-.125emX}}

\begin{document}

\title{On Preconditioning Electromagnetic Integral Equations in the High Frequency Regime via Helmholtz Operators and quasi-Helmholtz Projectors
\thanks{This work was supported by the European Research Council (ERC) under the European Union’s Horizon 2020 research and innovation program (grant agreement No 724846, project 321).}
}

\author{\IEEEauthorblockN{Alexandre Dély, Adrien Merlini, Simon B. Adrian, Francesco P. Andriulli}
\IEEEauthorblockA{Department of Electronics and Telecommunications, Politecnico di Torino, Turin, Italy \\
alexandre.dely@polito.it, adrien.merlini@polito.it, simon.adrian@polito.it, francesco.andriulli@polito.it}
}

\maketitle

\begin{abstract}

Fast and accurate resolution of electromagnetic problems via the \ac{BEM} is oftentimes challenged by conditioning issues occurring in three distinct regimes: (i) when the frequency decreases and the discretization density remains constant, (ii) when the frequency is kept constant while the discretization is refined and (iii) when the frequency increases along with the discretization density. While satisfactory remedies to the problems arising in regimes (i) and (ii), respectively based on Helmholtz decompositions and \calderon{}-like techniques have been presented, the last regime is still challenging. In fact, this last regime is plagued by both spurious resonances and ill-conditioning, the former can be tackled via combined field strategies and is not the topic of this work. In this contribution new symmetric scalar and vectorial electric type formulations that remain well-conditioned in all of the aforementioned regimes and that do not require barycentric discretization of the dense electromagnetic potential operators are presented along with a spherical harmonics analysis illustrating their key properties.

\end{abstract}

\begin{IEEEkeywords}
electric field integral equation, condition number, Helmholtz decomposition, high frequency simulation
\end{IEEEkeywords}

\acresetall

\section{Introduction}

The \ac{BEM} is one of the most widespread schemes for solving problems of electromagnetic scattering by \ac{PEC} objects. Its popularity stems from the relatively low number of unknowns that need to be solved for, because it only requires discretization of the surface of the scatterer, from its automatic enforcing of radiation condition and from its immunity to numerical dispersion. Since the system matrices obtained via \ac{BEM} are dense, fast algorithms have been introduced to linearize the complexity of the resolution process.

For handling complex simulation scenarios, the integral operators making up the electromagnetic formulation must be well-conditioned. However, most integral operators are ill-conditioned in at least one regime: (i) in the low frequency regime characterized by a fixed discretization and a decreasing frequency and (ii) in the dense discretization regime in which the frequency is kept constant while the discretization density increases. Several solutions have been proposed to stabilize the conditioning of standard electromagnetic integral equations in these two regimes. However, a third regime (iii) in which the frequency increases along with the discretization density also causes an unbounded increase of the condition number.

Remedies for problems (i) and (ii) of the \ac{EFIE} have been the focus of numerous studies. Its low frequency behavior is typically cured through an independent frequency re-scaling of the solenoidal and non-solenoidal parts of its solution permitted by a Helmholtz decomposition, such as Loop-Star/Loop-Tree, or through the computation of auxiliary variables. Both methods do however suffer from limitations, the former further degrades the dense discretization breakdown while the latter has a significant computational overhead. The dense discretization breakdown has also been extensively investigated and several of its cures leverage the \calderon{} identities that demonstrate that the \ac{EFIO} can potentially precondition itself. A new \ac{EFIE} formulation combining the \calderon{} identities and the quasi-Helmholtz projectors has more recently been introduced to simultaneously address issues (i) and (ii) \cite{andriulliWellConditionedElectricField2013}. An equivalent technique that does not require barycentric refinement of the discretized geometry has also been introduced \cite{adrian2019refinement}. However, neither of these techniques is able to address the high frequency breakdown (iii).

The high frequency breakdown should not be confused with the spurious internal resonance problem plaguing certain integral operators on closed structures in high frequency simulations. It is a well known issue which is traditionally cured by using a \ac{CFIE}. These resonances are not treated in this paper to focus specifically on the second problem which is an unbounded increase of the condition number (after removing the resonances) as the frequency increases along with the discretization density.

In this contribution two new formulations, one scalar and one vectorial, capable of handling problem (i), (ii) and (iii) on a large class of geometries are presented. Both schemes are symmetric and neither requires the discretization of the dense electromagnetic operators on the barycentric mesh.
The frequency regularization, both at low and high frequency, is performed by leveraging on the quasi-Helmholtz projectors and Helmholtz operators while the refinement-related ill-conditioning is treated using a \calderon-like scheme. The new formulations will be presented along with a spherical harmonics analysis confirming the theoretically predicted behavior of the new schemes.

\section{Notation and Background}

Given a simply connected \ac{PEC} object of surface $\Gamma$ living in a background medium of permitivity $\varepsilon$ and permeability $\mu$, the electric surface current density $\Jv$ induced on $\Gamma$ by a time-harmonic incident electric field $\Ev^\text{i}$ of frequency $f$ can be obtained by solving the \ac{EFIE}
\begin{equation}
    \eta  \left(\Top \Jv\right)(\vr) = -\n(\vr) \times \Ev^\text{i}(\vr)\,,  \label{eq:contefie}
\end{equation}
with
\begin{align}
\left(\Top\Jv\right)\left(\vr\right) & = -\jm k  \left(\Ts\Jv\right)\left(\vr\right) - \frac{1}{-\jm  k} \left(\Th\Jv\right)\left(\vr\right)\,,\\
\left(\Ts\Jv\right)\left(\vr\right) & = \n(\vr) \times \int_\Gamma \frac{\e^{-\jm k |\vr - \vrp|}}{4 \uppi |\vr - \vrp|} \Jv(\vrp) \dd S'\,,\\
\left(\Th\Jv\right)\left(\vr\right) & =  \n(\vr) \times \nabla \int_\Gamma \frac{\e^{-\jm k |\vr - \vrp| }}{4 \uppi |\vr - \vrp|} \nabla' \cdot \Jv(\vrp) \dd S'\,,
\end{align}
where $\eta=\sqrt{\mu / \varepsilon}$, $k = 2 \uppi f \sqrt{\mu \varepsilon}$ and $\uv{n}$ is the normal to $\Gamma$. This equation can be numerically solved via the \ac{BEM} by first expanding the unknown as a linear combination of $N$ \ac{RWG} basis functions $\{\RWG_i\}$ ($\veg j = \sum_{i=1}^N \left[ \vec j \right]_i \RWG_i$) and forming a system matrix by testing the resulting discretized equation with rotated \ac{RWG} functions $\{\nxRWG_i\}$
\begin{equation}
    \eta \mat T \vec j = \vec e^\mathrm{i}\,,
\end{equation}
where $\left[\vec e^\mathrm{i}\right]_{i} = \left< \nxRWG_i, -\uv{n} \times \Ev^\text{i} \right>$, $\mat T = -\jm k \mat T_s - (-\jm k)^{-1} \mat T_h$, $\left[\matTs \right]_{ij} = \left< \nxRWG_i\,, \Ts \RWG_j \right>$ and $\left[\matTh \right]_{ij} = \left< \nxRWG_i\,, \Th \RWG_j \right>$.
In addition to these discretized operators we denote the Gram matrices and mix-Gram matrices as
\begin{align}
    \left[\mat{G}_{a b} \right]_{ij} &= \left< a_i\,, b_j \right>\,,
\end{align}
where $a$ and $b$ can be any valid combination of the following basis functions: the \ac{RWG} $\RWG$ and \ac{BC} $\BC$ \cite{buffa2007dual} basis functions, the rotated \ac{RWG} $\uv{n} \times \RWG$ functions, the pyramid $\Pyramid$ and dual pyramid $\DualPyramid$ basis functions and the patch $\Patch$ and dual patch $\DualPatch$ basis functions. A detailed definition of these functions can be found in \cite{buffa2007dual}.
%

\section{New High Frequency Stable Electric Type Equations}

In the contribution we present two different formulations:  one adapted for the preconditioning of the \ac{EFIE} after a Loop-Star decomposition, the other suited for preconditioning directly the un-decomposed \ac{EFIE} operator.

\subsection{Scalar Formulation}

In the new scalar formulation, the Loop and the Star components of the $\uv{n} \times \Top$ operator
\begin{align}
   \Top_\Lambda &= \nabla \cdot  \uv{n} \times \left( \uv{n} \times \Top \right) \uv{n} \times \nabla   \,, \\
    \Top_\Sigma &=  \laplacian{}^{-1} \nabla \cdot  \left( \uv{n} \times \Top \right) \nabla  \laplacian{}^{-1} \,,
\end{align}
are independently preconditioned with the corresponding scalar Helmholtz operators to form Loop and Star blocks that are immune from breakdowns (i) to (iii)
\begin{gather}
 \label{eq:stable_scalar_operator_loop}
    k^{-2} \Top_\Lambda \laplacian{}^{-1} \left( \laplacian{} + \km^2 \Iop \right) \laplacian{}^{-1} \Top_\Lambda \,,\\
    \label{eq:stable_scalar_operator_star}
    k^{2} \Top_\Sigma \laplacian{} \left( \laplacian{} + \km^2 \Iop \right)^{-1} \laplacian{} \Top_\Sigma
  \,,
\end{gather}
where $\laplacian{}$ denotes the Laplace-Beltrami operator on $\Gamma$ and $\km=k+0.4\jm k^{1/3} R^{-2/3}$ is a modified wave number capable of stabilizing the norm of $\op T$ once multiplied with the Laplace-Beltrami operator on a sphere of radius $R$ \cite{darbas2015well}.
These block operators can then be discretized in a Galerkin setting to form the stable discretized block operator 
\begin{equation}
        \label{eq:scalar_formulation}
    \mat Z^s =  \mat T_{\LS}^\T \matel{ \mat{L}_{\text{L}}^\pinv{} \mat{H}_{\text{L}} \mat{L}_{\text{L}}^\pinv{} & \matO \\
     \matO & \mat{G}_{\DualPyramid \Patch}^{-1} \mat{L}_{\text{S}} \mat H_{\text{S}}^\pinv{} \mat{L}_{\text{S}} \mat{G}_{\Patch \DualPyramid }^{-1} }  \mat T_{\LS}  \,,
\end{equation}
where
\begin{gather}
    \mat{L}_{\text{L}} =  - \matL^\T \mat{G}_{\RWG \RWG} \matL  \,, \\
    \mat{L}_{\text{S}} = - \matS^\T \mat{G}_{\BC \BC} \matS \,, \\
    \mat H_{\text{L}} = \mat{L}_{\text{L}} + \km^2 \gramPyPy  \,, \\
    \mat H_{\text{S}} = \mat{L}_{\text{S}} + \km^2 \gramDPyDPy  \,, \\
    \widetilde{\matS} = \matS \left(\matS^\T \matS \right)^\pinv{} \mat{G}_{\Patch \Patch} \,,\\
    \mat T_{\LS} = \matel{(-\jm k)^{-1} \matL^\T \\
                            \widetilde{\matS}^\T} \mat T \matel{\matL & - \jm k \widetilde{\matS}} \,.
\end{gather}
In addition, for this operator to be stable until arbitrarily low frequencies the terms $\matL^\T \matTh$ and $\matTh \matL$ must be explicitly set to $\matO$.
The preconditioned system \eqref{eq:scalar_formulation} still exhibits a nullspace of dimension $2$ that corresponds to the all one vectors that are in the nullspaces of $\matL$ and $\matS$, both of which can be removed by deflection but these passages are not detailed here for the sake of brevity.

\subsection{Vectorial Formulation}

Along with the new scalar formulation we present a new vector formulation that leverages on the quasi-Helmholtz projectors $\Pl = \matL ( \matL^\T \matL ) \matL^\T$ and $\Ps = \matS ( \matS^\T \matS ) \matS^\T$ instead of Loop-Star techniques to perform the Helmholtz decomposition of $\uv{n} \times \Top$. The decomposed operators are then preconditioned using vector Helmholtz operators, which for the solenoidal operator yields
\begin{align}
\label{eq:stable_vector_operator_solenoidal}
    k^{-2} \uv{n} \times \Top \left(\veclaplacian{} + \km^2 \Iop \right) \uv{n} \times \Top \,,
\end{align}
and for the non-solenoidal operator
\begin{align}
\label{eq:stable_vector_operator_nonsolenoidal}
    k^2 \uv{n} \times \Top \left(\veclaplacian{} + \km^2 \Iop \right)^{-1} \uv{n} \times \Top \,,
\end{align}
where $\veclaplacian{}$ is the vector Laplacian on $\Gamma$.
The complete, stabilized, operator can then be discretized as
\begin{equation}
    \mat Z^v = \mat{T}_{\matL \matS}^\T \left( \mat{G}_{\BC , \uv{n} \times \RWG}^{-1}  \mat H_{\matL} \mat{G}_{\uv{n} \times \RWG, \BC}^{-1} + \mat H_{\matS}^\pinv{} \right) \mat{T}_{\matL \matS} \,,
\end{equation}
where
\begin{align}
    \widetilde{\mat{L}}_{\matL} &= \mat{G}_{\Pyramid \DualPatch}^{-1} \mat{G}_{\Pyramid \Pyramid}  \left( \matL^\T \mat{G}_{\RWG \RWG} \matL \right)^\pinv{} \mat{G}_{\Pyramid \Pyramid}  \mat{G}_{\DualPatch \Pyramid}^{-1} \,,\\
    \widetilde{\mat{L}}_{\matS} &= \mat{G}_{\DualPyramid \Patch}^{-1} \mat{G}_{\DualPyramid \DualPyramid} \left( \matS^\T \mat{G}_{\BC \BC} \matS \right)^\pinv{} \mat{G}_{\DualPyramid \DualPyramid}  \mat{G}_{\Patch \DualPyramid}^{-1} \,, \\
    \mat H_{\matL} &= \matL \left( - \mat{G}_{\DualPatch \DualPatch}^{-1}  + \km^2 \widetilde{\mat{L}}_{\matL} \right) \matL^\T  \,, \\
    \mat H_{\matS} &= \matS \left( - \gramPaPa^{-1} + \km^2  \widetilde{\mat{L}}_{\matS} \right) \matS^\T \,, \\
    \mat{T}_{\matL \matS} &= \left((-\jm k)^{-1} \Pl + \Ps \right) \mat{T} \left( \Pl - \jm k \Ps \right)  \,.
\end{align}
As for the scalar case, the low frequency stability of the formulation is dependent on the explicit cancellation of the terms $\Pl \matTh$, $\matTh \Pl $, $\Pl \mat H_{\matS}^\pinv{}$, $\mat H_{\matS}^\pinv{} \Pl$, $\Ps \mat{G}_{\BC, \uv{n} \times \RWG}^{-1} \mat H_{\matL}$ and $\mat H_{\matL} \mat{G}_{\uv{n} \times \RWG, \BC}^{-1} \Ps$. We omit the passages for the sake of brevity.

\section{Numerical Results}

To verify the stability of the newly introduced formulations we perform a spherical harmonics analysis of the continuous operators in the case of
a sphere of radius $R$ for which we denote as $Y_{lm}$ the spherical harmonic of order $l$ ($m \in [-l,l]$). 
It can be demonstrated that $Y_{lm}$ and $\uv{n} \times \nabla Y_{lm}$ are respectively the eigenvectors of \eqref{eq:stable_scalar_operator_loop} and \eqref{eq:stable_vector_operator_solenoidal} which also share the same eigenvalues
\begin{equation}
    \sigma_{\mat{\Lambda}}(l,k) = k^{-2} \left( \riccatibessel{l}(k R) \riccatihankeltwo{l}(k R) \right)^2 \left( - \frac{l (l+1)}{R^2} + \km^2 \right)\,,
\end{equation}
where the subscript $m$ has been omitted because the eigenvalues are identical  for each $m \in [-l,l]$ (they have multiplicity $2l+1$) and
where $\riccatibessel{l}$ and $\riccatihankeltwo{l}$ are the Riccati-Bessel and Riccati-Hankel functions.
Similarly, $Y_{lm}$ and $\nabla Y_{lm}$ are respectively the eigenvectors of \eqref{eq:stable_scalar_operator_star} and \eqref{eq:stable_vector_operator_nonsolenoidal} associated to the shared eigenvalues
\begin{align}
    &\sigma_{\mat{\Sigma}}(l,k)  = k^{2} \left( {\riccatibessel{l}}'(k R) {\riccatihankeltwo{l}}'(k R) \right)^2 \left( - \frac{l (l+1)}{R^2} + \km^2 \right)^{-1}\,,
\end{align}
where ${}'$ denotes the derivative.

Careful analysis of these eigenvalues  shows that the regularized operators are both dense discretization and high-frequency stable (\Cref{fig:SH_stable_operatorL,fig:SH_stable_operatorS}). The dense discretization stability is characterized by the clustering of the absolute value of the eigenvalues at $0.25$, while the high frequency stability is a consequence of the fact that their absolute value remains bounded by $1$. These operators will hence be high-frequency stable after eliminating their spurious resonances.
Finally, the formulation is also low frequency stable since both $\sigma_{\mat{\Lambda}}(l,k)$ and $\sigma_{\mat{\Sigma}}(l,k)$ converge to a finite non-zero value ($l(l+1)/(2l+1)^2$) that is independent of the frequency as $k \to 0$.

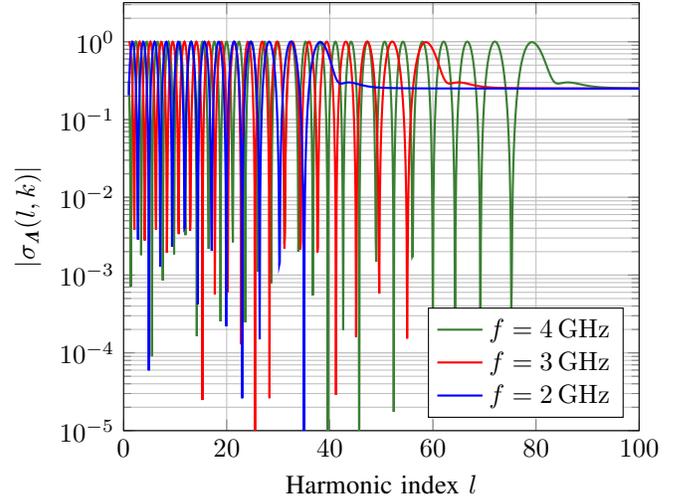
\begin{figure}
    \centering
        \begin{tikzpicture}[scale=1.0]
        \begin{axis}[
            grid=both,
            xlabel={Harmonic index $l$},
            ylabel={$\left| \sigma_{\mat{\Lambda}}(l,k) \right|$},
            ymode=log,
            legend pos=south east,
            xmin=0, xmax=100,ymin=1e-5
                ]
        \addplot+[color=OliveGreen,mark=none,thick] table [x=index, y=evL3]{figures/data/SH_stable_operators.txt};
        \addlegendentry{$f=\SI{4}{\giga\hertz}$};
        \addplot+[color=red,mark=none,thick] table [x=index, y=evL2]{figures/data/SH_stable_operators.txt};
        \addlegendentry{$f=\SI{3}{\giga\hertz}$};
        \addplot+[color=blue,mark=none,thick] table [x=index, y=evL1]{figures/data/SH_stable_operators.txt};
        \addlegendentry{$f=\SI{2}{\giga\hertz}$};
        \end{axis}
    \end{tikzpicture}
    \caption{Absolute value of the eigenvalues of the high frequency stable operators (solenoidal/loop) for different frequencies, which show a bound at $1$ and a clustering at $0.25$ for large $l$.}
    \label{fig:SH_stable_operatorL}
\end{figure}
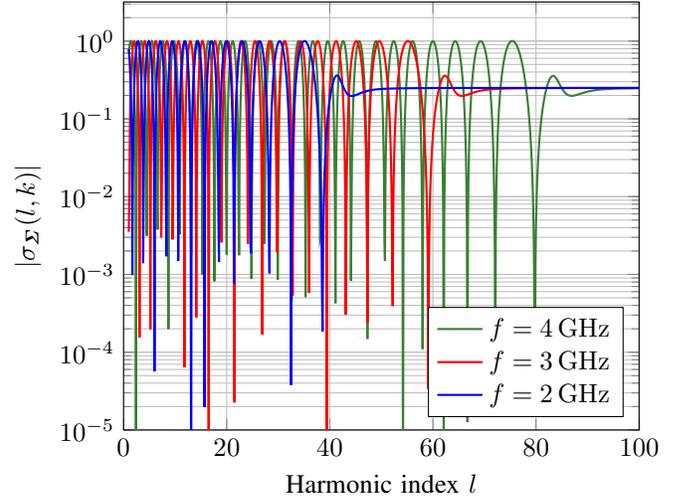
\begin{figure}
    \centering
        \begin{tikzpicture}[scale=1.0]
        \begin{axis}[
            grid=both,
            xlabel={Harmonic index $l$},
            ylabel={$\left| \sigma_{\mat{\Sigma}}(l,k) \right|$},
            ymode=log,
            legend pos=south east,
            xmin=0, xmax=100,ymin=1e-5
                ]
        \addplot+[color=OliveGreen,mark=none,thick] table [x=index, y=evS3]{figures/data/SH_stable_operators.txt};
        \addlegendentry{$f=\SI{4}{\giga\hertz}$};
        \addplot+[color=red,mark=none,thick] table [x=index, y=evS2]{figures/data/SH_stable_operators.txt};
        \addlegendentry{$f=\SI{3}{\giga\hertz}$};
        \addplot+[color=blue,mark=none,thick] table [x=index, y=evS1]{figures/data/SH_stable_operators.txt};
        \addlegendentry{$f=\SI{2}{\giga\hertz}$};
        \end{axis}
    \end{tikzpicture}
    \caption{Absolute value of the eigenvalues of the high frequency stable operators (non-solenoidal/star) for different frequencies, which show a bound at $1$ and a clustering at $0.25$ for large $l$.}
    \label{fig:SH_stable_operatorS}
\end{figure}

\section{Conclusion}

We have presented two new electric integral formulations that are stable in the high frequency regime (resonances excluded) in addition to being stable in the low frequency and dense discretization regimes. These new formulations are stabilized using Helmholtz operators with modified wavenumbers: the first one is based on a Loop-Star decomposition and leverages the scalar Helmholtz operator while the second one is based on the quasi-Helmholtz projectors and leverages a vector Helmholtz operator. In both case, the resulting preconditioned system is symmetric and does not require the use of dual functions for the discretization of the electromagnetic potential operators.

\bibliographystyle{IEEEtran}
\bibliography{IEEEabrv,ref}

\end{document}